\documentclass[pre, aps,showpacs,nofootinbib, onecolumn]{revtex4-1}

	\pdfoutput=1
	
	\usepackage{amsmath,amssymb,amsfonts,mathtools}
	\usepackage{graphicx}
	\usepackage{placeins}
	\usepackage{upgreek}
	\usepackage{xcolor}
	\usepackage{standalone}
	\usepackage{tikz}
	\usetikzlibrary{positioning, patterns, calc, shapes.misc, 3d, shapes.arrows}
	\usepackage{xstring}
	\usepackage{multirow}
	\usepackage{tabularx}
	\usepackage{ragged2e}

	\graphicspath{{./},{./images/}}

    \newcolumntype{C}[1]{>{\centering\arraybackslash}p{#1}}

	\newcommand{\cref}[1]{Ref.~\cite{#1}}

	\newcommand{\bea}{\begin{eqnarray}}
	\newcommand{\eea}{\end{eqnarray}}
	\newcommand{\vect}[1]{\mathbf{#1}}

	\newcommand{\Fid}{\mathcal{F}^\text{id}}
	\newcommand{\Fex}{\mathcal{F}^\text{ex}}

	\newcommand{\tFAO}{\tilde{\mathcal{F}}_\text{AO}}
	\newcommand{\Flg}{\mathcal{F}_\text{lg}}

	\newcommand{\kt}{k_\text{B}T}

	\newcommand{\rhoc}{\rho_\text{c}}
	\newcommand{\req}{\rho_\text{c,eq}}
	\newcommand{\sums}{\sum_{\pmb{s}}}
	\newcommand{\Ve}{V_\text{ext}}
	\newcommand{\pzd}{\Phi^\text{0D}}

	\newcommand{\mucoex}{\mu_{\text{coex}}}

	\newcommand{\rhopr}{\rho^\text{res}_{\text{p}} }
	\newcommand{\rhopcan}{\rho_{\text{pc}^n,\alpha}  }
	\newcommand{\rhopca}{\rho_{\text{pc},\alpha}  }
	\newcommand{\rhopcaeq}{\rho_{\text{pc-eq},\alpha}  }

	\newcommand{\rholc}{\rho_{\text{l,coex}}}
	\newcommand{\rhovc}{\rho_{\text{v,coex}}}
	\newcommand{\rl}{\rho_{\text{l}}}
	\newcommand{\rv}{\rho_{\text{v}}}
	\newcommand{\pli}{p_{\text{l}}}
	\newcommand{\pv}{p_{\text{v}}}

	\newcommand{\dF}{\Delta F}
	\newcommand{\vd}{V_\text{d}}
	\newcommand{\ad}{A_\text{d}}
	\newcommand{\rd}{R_\text{d}}
	\newcommand{\Rem}{R_\text{e}}
	\newcommand{\Rs}{R_\text{s}}
	\newcommand{\Vd}{{\cal V}_\text{d}}
	\newcommand{\Ad}{{\cal A}_\text{d}}
	\newcommand{\tg}{\tilde\gamma_\infty}

\begin{document}

	\title{Droplet condensation in the lattice gas with density functional theory}
	\author{M.~Maeritz and M. Oettel}
	\email{Email address: martin.oettel@uni-tuebingen.de}
	\affiliation{Institut f\"ur Angewandte Physik, Eberhard Karls Universit\"at T\"ubingen, Auf der Morgenstelle 10, D-72076 T\"ubingen, Germany}

	\begin{abstract}
A density functional for the lattice gas (Ising model) from fundamental measure theory is applied to the problem of droplet states in three--dimensional, finite systems. Similar to previous simulation studies, the sequence of droplets changing to cylinders and to planar slabs  is found   upon increasing the average density $\bar\rho$ in the system. Owing to the discreteness of the lattice, additional effects in the state curve for the chemical potential $\mu(\bar\rho)$ are seen upon lowering the temperature away from the critical temperature (oscillations in $\mu(\bar\rho)$ in the slab portion and spiky undulations in $\mu(\bar\rho)$ in the cylinder portion as well as an undulatory behavior of
the radius of the surface of tension $\Rs$ in the droplet region). This behavior in the cylinder and droplet region is related to washed--out layering transitions at the surface of liquid cylinders and droplets. The analysis of the large--radius behavior of the surface tension $\gamma(\Rs)$ gave a dominant contribution $\propto 1/\Rs^2$, although the consistency of $\gamma(\Rs)$ with the asymptotic behavior of the radius--dependent Tolman length seems to suggest a weak logarithmic contribution $\propto \ln\Rs/\Rs^2$ in $\gamma(\Rs)$. The coefficient of this logarithmic term is smaller than a universal value derived with field--theoretic methods.
	\end{abstract}

	\maketitle

	\section{Introduction}

 	For the practical description of phase transitions in the nucleation regime, the free energy of a critical nucleus of the newly forming phase plays a prominent role. This is manifest 
	in classical nucleation theory (CNT)  for liquid drop nucleation in a supersaturated gas 
	where the formation free energy $\dF(\vd)$ is simply described by a sum
	of a volume term (proportional to the droplet volume $\vd$ in a supersaturated bulk state) and 
	a surface term \cite{Kashchiev2000}. It is essential in kinetic approaches  
	where a formation rate of new droplets is proportional to $\exp( -\beta \dF(\vd))$ and thus inverse exponentially
	dependent on the formation free energy  (here, $\beta=1/(\kt)$ is the inverse 
	temperature). However, a critical droplet is at best metastable and it is problematic to associate a free energy 
	with it since the free energy is an equilibrium concept. 

	In order to avoid such complications, a thermodynamic method for the estimation of $\dF(\vd)$ has been developed 
	some years ago \cite{Schrader2009,Block2010,Troester2012,Binder2012}. It is based on the analysis of equilibrium states in a finite volume (box) (with
	periodic boundary conditions) in the canonical
	ensemble. For a given box size in three dimensions (3D), the variation of the particle number (density) 
	from the coexistence vapor density
	to the coexistence liquid density leads through a sequence of states described by 
	\textit{supersaturated vapor $\to$ droplets $\to$ liquid cylinders $\to$ planar interfaces $\to$ vapor cylinders
	$\to$ bubbles $\to$ undersaturated liquid}. The excess free energy over the supersaturated state (formation free energy
	of droplets and cylinders) can be unambiguously computed since it belongs to an equilibrium state. In a box with fixed size,
	the range of droplet radii in the droplet state is limited, and one obtains results for a larger range of droplet radii
	by combining the results for different box sizes. 

	This thermodynamic method has been mainly advanced with simulations. The extraction of a radius--dependent surface tension
	for droplets has been shown to be influenced by strong finite--size effects whose analysis leads to the conclusion
	that an asymptotic expansion of the surface tension in the inverse radius $1/\rd$ should not only contain powers of $1/\rd$  
	but also terms logarithmic in $\rd$. In 3D, the leading term, however, is the term $\propto 1/\rd$ containing the
	famous ``Tolman length'' but the next term is of this logarithmic type and  varies $\propto \ln \rd / \rd^2$ \cite{Wallace1980}.
	The available results in the literature show that the extraction from simulations of the Tolman length and other, sub--leading terms 
	in the expansion of the surface tension is by no means an easy task \cite{Schrader2009,Troester2017,Song2019}. 
	This gives a motivation to use
	theoretical tools in applying the thermodynamic method, notably density functional theory (DFT). 
	For most systems of interest which show liquid-vapor
	phase transitions the free energy functional (as the necessary prerequisite) is known only approximately but in contrast to
	simulations the extrapolation of droplet surface tensions to large radii is easier. 
	A previously studied example was the Lennard--Jones fluid where
	the simulation results indicate a convergence of the asymptotic Tolman length to the DFT result \cite{Block2010}. 
	However, the next, subleading term did not agree at all. Most likely this was due to the neglect of translational
	entropy of the droplet in the analysis of the simulation results (such a term is absent in DFT calculations where the droplet
	is always in the center of the numerical box).

    From a computational point of view lattice models are easier, both for the elucidation
    of phase transition and wetting phenomena  with DFT \cite{Archer2014,Archer2017} and for simulations.
	The comparable ease of simulations has triggered many studies of droplets in the lattice gas or Ising model. Among these are studies focussing on establishing the droplet condensation/evaporation transition, see e.g. Refs.~\cite{Biskup2003,Nussbaumer2006} and others looking more in detail on the actual nucleation process \cite{Stauffer1982,Ryu2010,Schmitz2013}. 
	Owing to the computational advantage, the problem of the leading and
	subleading contributions to the expansion of the surface tension (including the problem of the translational entropy)
	was revisited in Ref.~\cite{Troester2017} in the context of lattice models, the Ising model (lattice gas) in 2D and an Ising--type model 
	with three--spin interactions in 3D. Therefore it appears to be a worthwhile exercise to apply DFT methods to the droplet problem
	in lattice models for complementary results on the expansion of the surface tension, using also the same type of finite boxes (note that in DFT studies of continuum models usually spherical boxes are employed \cite{Koga1998,Block2010,Gross2018}).
	In the present study we use a recently developed density functional for the lattice gas
	 \cite{Maeritz2021} (much improved compared to the standard mean--field functional) for the description of droplet states in finite
	cubic boxes. We find the sequence of states as described above but we also find certain effects of the lattice discreteness 
	in these states which had not been reported so far. These include the existence of equilibrium, planar interfaces under
	tension at chemical potentials off coexistence (described in detail in Ref.~\cite{Maeritz2021}) and certain layering effects for the droplet and cylinder states.
	The symmetry of the lattice gas shows that the expansion of the surface tension must involve a zero Tolman length,
	so the leading coefficients in 3D show a dependence $\propto \ln \rd / \rd^2$ and $\propto 1 /\rd^2$.

	The paper is structured as follows. In Sec.~\ref{sec:theory} we recapitulate the essence of the lattice gas functional derived in Ref.~\cite{Maeritz2021} as well as the phenomenological theory of droplet states. Sec.~\ref{sec:results} shows results for the possible states inside the gas--liquid binodal at three selected temperatures and discusses the asymptotic expansions of the droplet surface tension and a radius--dependent Tolman length. In Sec.~\ref{sec:summary} we summarize and conclude our work.

	\section{Theory}
\label{sec:theory}

    \subsection{Density functional theory for the lattice gas}

	The lattice gas model on a simple cubic lattice is defined 
	in terms of hard particles with mutual exclusion on the same lattice site and which have nearest--neighbor attractions of strength $-\epsilon$.
	It is equivalent to the Ising model on the respective lattices.  

    We define an ensemble--averaged density $\rhoc(\pmb{s})$ for lattice gas particles (where the index ``c'' stands for colloid)
	on discrete lattice sites $\pmb{s}$ of the lattice.
    In DFT, one defines a functional for the grand potential \cite{Evans1979}
	\bea
	  \Omega[\rhoc] = \Fid[\rhoc] + \Fex[\rhoc] - \sums (\mu - \Ve(\pmb{s})) \rhoc(\pmb{s})
	\eea
	which is split into an ideal gas free energy functional $\Fid$, an excess free energy functional $\Fex$ and a remaining part
	containing the chemical potential $\mu$ and the contribution of a one--body external potential $\Ve(\pmb{s})$. The ideal
	gas free energy functional is given by
	\bea
	  \beta \Fid[\rhoc] = \sums \rhoc(\pmb{s}) ( \ln \rhoc(\pmb{s}) -1)
	  \label{eq:fid}
	\eea
	(where $\beta= 1/(\kt)$ is the inverse temperature) and the excess functional is generally unknown.
    
    In Ref.~\cite{Maeritz2021}, we have derived an approximate excess functional through the following procedure. First, the attractions
    between the particles are viewed as effective depletion interactions between the particles and lattice polymers of length 2 with which the particles interact hard. In 3D, there are 3 species of such polymers and correspond to the orientation $\alpha \in \{x,y,z\}$
    of the polymer particle on the cubic lattice. The lattice polymers do not interact with each other and hence form a three--component
    ideal gas. (In spirit, this is a lattice Asakura--Oosawa (AO) model.) Second,
    the partition function of the lattice polymers is reformulated in terms of a partition function for a mixture of polymer clusters,
    distinguished by their orientation $\alpha$ and the number $n$ of polymers they contain. Two clusters of polymers with the same $\alpha$
    but different $n$ cannot occupy the same lattice site, since the simultaneous presence of two clusters with size $n$ and $m$ corresponds to a cluster of size $n+m$. Therefore one can represent the free energy functional of the polymers in terms
    of the functional for a multicomponent hard lattice gas of clusters. 
    Third, the excess free energy functional of the interacting system
    of lattice gas particles and polymer clusters is approximated using lattice fundamental measure theory (FMT).
    
    The results of Ref.~\cite{Maeritz2021} are summarized best by first introducing an auxiliary functional $\tilde \varUpsilon$ 
    for the lattice AO system, i.e. the system of lattice gas 
    particles with density $\rhoc(\pmb{s})$ coupled grand--canonically to the polymer clusters with orientation $\alpha$ and with a total density $\rhopca(\pmb{s})=\sum_n \rhopcan(\pmb{s})$ where $\rhopcan(\pmb{s})$ is the density of clusters of size $n$ and
    orientation $\alpha$. The polymer cluster chemical potential $\beta^{-1} \ln \zeta_\alpha$ is the same for all orientations, which defines an orientation--independent activity $\zeta_\alpha= \zeta$:
    \bea
	  \beta \tilde \varUpsilon[\rho_\text{c}; \{ \rhopca \} ] &=& 
                      \beta \Fid[\rhoc] + \sum_\alpha \beta \Fid[\rhopca]
		      + \beta\tFAO^\text{ex}[\rhoc, \{\rhopca\}] \nonumber \\
		       & & - \ln \zeta   \sums \sum_\alpha \rhopca(\pmb{s}) \;.  \label{eq:Upst}
    \eea
    The activity of the polymer clusters is related to the attraction strength $\epsilon$ by
    \bea
       \zeta = \exp(\beta\epsilon) - 1 \;.
    \eea
    The lattice FMT expression for the excess free energy $\tFAO^\text{ex}[\rhoc, \{\rhopca\}] $ of lattice gas particles and 
    polymer clusters is given in App.~\ref{app:fex}.
    
    Minimization of this lattice AO auxiliary functional with respect to $\rhopca(\pmb{s})$ gives an effective AO functional,
    depending only on the lattice gas particle density profile $\rhoc(\pmb{s})$ and the attraction strength through $\zeta$: 
    \bea
	  \tFAO^\text{eff}[\rhoc(\pmb{s});\epsilon(\zeta)] &=& \text{min}_{\rhopca(\pmb{s})} \tilde \varUpsilon[\rho_\text{c}; \{ \rhopca \} ]
    \eea
  	Finally, the full lattice gas functional is obtained by performing a necessary subtraction of a constant and one--body term
  	in $\rhoc(\pmb{s})$ and is given by:
	\bea
         \Flg[\rhoc(\pmb{s}); \epsilon(\zeta)] = \tFAO^\text{eff}[\rhoc(\pmb{s}); \rhopr] + \rhopr \sums \sum_\alpha
                                           \left( 1 - \sum_{\pmb{s}' \in \{\pmb{s},\pmb{s}+\hat{\pmb{e}}_\alpha\}} \rhoc(\pmb{s}') \right) \;. 
     \label{eq:flg_final}
    \eea  
    Here, $\hat{\pmb{e}}_\alpha$ is a unit vector on the lattice in $\alpha$--direction and $\rhopr=\beta\epsilon=\ln(\zeta+1)$ is the reservoir density of polymers (which is the same for all polymer species).  
    We name this functional the ``Highlander functional'', following the nomenclature in Ref.~\cite{Cuesta2005} where the idea of using polymer clusters in the treatment of lattice AO models was introduced first.

    Note that this functional is superior to the standard mean--field functional, defined by 
    \bea
	  \Flg^\text{mf}[\rhoc] = \beta^{-1} \sums \left( \rhoc(\pmb{s}) \ln \rhoc(\pmb{s})  + [1-\rhoc(\pmb{s})] \ln[1-\rhoc(\pmb{s})] \right) - \frac{\epsilon}{2} \sums \sum_{\pmb{s}' \in \text{n.n.}(\pmb{s})} \rhoc(\pmb{s}) \rhoc(\pmb{s}') \;.  
	 \label{eq:fmf}
	\eea
	Here, the first term is the exact functional for the hard lattice gas and the second term gives the
	contribution from attractions in random phase approximation \cite{Evans1979}.
	The summation over lattice points $\pmb{s}'$ is restricted to nearest neighbors of $\pmb{s}$ ($\text{n.n.}(\pmb{s})$). For bulk fluids, the standard mean--field functional gives a phase diagram like the Bragg--Williams approximation and the Highlander functional results in the more precise Bethe--Peierls approximation for that. Planar surface tensions in 3D evaluated with the Highlander functional are close to simulation results \cite{Maeritz2021}.  

    \subsection{Numerical minimization of the  Highlander functional}
    
    In order to obtain density profile solutions for droplets in finite boxes, one needs
    to perform the numerical minimization in 3D of $\Omega = \Flg - \sums \rhoc (\mu -\Ve)$ (with $\Flg$ given in
Eq.~(\ref{eq:flg_final})) with respect to $\rhoc(\pmb{s})$. However, the analytic derivative with respect to $\rhoc(\pmb{s})$
is quite involved, owing to the dependency of $\rhopca(\pmb{s})$ on $\rhoc(\pmb{s})$. It is therefore advisable
to minimize the  total particle--polymer grand functional
\bea
  \tilde \Omega [\rhoc, \{\rhopca\}] = \tilde \Upsilon - \sums \rhoc(\pmb{s}) (\mu -\Ve(\pmb{s})) {+ \rhopr \sums \sum_\alpha 
					   \left( 1- \sum_{\pmb{s}' \in \{\pmb{s},\pmb{s}+\hat{\pmb{e}}_\alpha\}} \rhoc(\pmb{s}') \right)}
\eea
(with $\tilde \Upsilon$ given in Eq.~(\ref{eq:Upst})) with respect to $\rhopca(\pmb{s})$ and $\rhoc(\pmb{s})$
simultaneously. The self--consistent equations for the colloid and polymer cluster density profiles
take a form suitable for Picard iteration. They read
\begin{eqnarray}
\rho_\text{c}(s) &=& z(\pmb{s})\, e^{6\beta\epsilon}\,
\frac{\splitfrac{\left(1-m_1(\pmb{s})\right)\left(1-m_2(\pmb{s}-\hat{\pmb{e}}_x)\right) \left( 1-m_3(\pmb{s})\right)\left( 1-m_4(\pmb{s}-\hat{\pmb{e}}_y)\right)}{\cdot \left(1-m_5(\pmb{s})\right)\left( 1-m_6(\pmb{s}-\hat{\pmb{e}}_z)\right)}}{\left(1-m_{10}(\pmb{s})\right)^{5}} \notag\\
\rho_{\text{pc},x}(s) &=& \zeta\;\frac{\left( 1-m_1(\pmb{s})\right)\left(1-m_2(\pmb{s})\right) }{\left(1-m_7(\pmb{s})\right)} \notag\\
\rho_{\text{pc},y}(s) &=& \zeta\;\frac{\left( 1-m_3(\pmb{s})\right)\left(1-m_4(\pmb{s})\right) }{\left(1-m_8(\pmb{s})\right)} \notag\\
\rho_{\text{pc},z}(s) &=& \zeta\;\frac{\left( 1-m_5(\pmb{s})\right)\left(1-m_6(\pmb{s})\right) }{\left(1-m_9(\pmb{s})\right)}\qquad.
\label{eq:min3d}
\end{eqnarray}
Here, $z(\pmb{s})=\exp(\beta[\mu -\Ve(\pmb{s})])$. The definitions of the weighted densities  $m_i(\pmb{s})$ is given in Eq.~(\ref{eq:n3D}). The Picard iterations are done in a standard manner with suitable mixing of old and new density profiles.

For the problem of inhomogeneous states (droplet, cylinder, slab) in the finite box, there is no external potential: $\Ve(\pmb{s})=0$.
For an initial state at a specific density, an excess density over the bulk gas density in an approximately spherical, cylindrical or slablike domain is chosen,
or a density profile from a previous iteration at a slightly different density is used.  
The final results are the equilibrated inhomogeneous density profiles $\req(\pmb{s})$ and the associated auxiliary
polymer cluster profile $\rhopcaeq(\pmb{s})$ as well as the total grand potential 
$\Omega=\tilde \Omega [ \req, \{ \rhopcaeq \}]$. There are hysteresis  effects in the system, i.e. at a specific average density near transition regions,
both a droplet resp. cylinder profile can be found in the droplet--cylinder transition region, or a cylinder resp. slab profile in the cylinder--slab transition region. Reported below are the profiles with the lowest free energy at a specific average density. From droplet profiles surface tensions and droplet radii are extracted
using the phenomenological considerations of the next section.

	\subsection{Phenomenological theory for the surface tension of droplets}

The treatment here is quite standard, see e.g. Ref.~\cite{Troester2012}. 
	We assume a stable state of a droplet in the center of a finite box $\cal V$ with volume $V$ at chemical potential $\mu$. 
	The chemical potential is not equal to the chemical potential at liquid--vapor coexistence, $\mu \neq \mucoex$.
	For a homogeneous system, the function $\rho(\mu)$ has two solutions near coexistence: for $\mu=\mucoex$ these
	are the coexistence densities $\rhovc$ (vapor) and $\rholc$ (liquid). For $\mu \neq \mucoex$ we associate with $\rv=\rho(\mu)$  
	the solution near $\rhovc$ and with  $\rl=\rho(\mu)$ the solution near $\rholc$. Similarly we define
	$\pv=p(\mu)$ and $\pli=p(\mu)$ where $p(\mu)$ describes the pressure as function of chemical potential for a bulk system.
	The droplet has a total free energy $F$ resp. a grand potential $\Omega$ and 
	is characterized by an equilibrium density profile $\rho(\vect r)$. The association of a droplet surface corresponding
	to this density profile is arbitrary but two definitions of a surface $\Ad = \partial \Vd$ 
	(where $\Vd$ is the spherical droplet with volume $\vd$) are common and useful. The first is the notion
	of an equimolar surface defined by 
	\bea
           \int_{\Vd} (\rl - \rho(\vect r)) d\vect r = \int_{{\cal V} \backslash \Vd} (\rho(\vect r)-\rv) d\vect r \;. 
	\eea
	The condition is equivalent to the statement that the
        total volume can be divided into a homogeneous liquid inside $\Vd$ with density $\rl$ and a homogeneous
	vapor in ${\cal V} \backslash \Vd$ with density $\rv$ such that the
	total number of particles $N= \int_{\cal V}\rho(\vect r)d\vect r=\rl \vd + \rv(V-\vd)$ (no excess adsorption
	at the droplet surface). 
	The drop is assumed to be spherical with radius $\Rem$ (equimolar radius) which follows from the above condition
	as
	\bea
	  \Rem = \sqrt[3]{\frac{3}{4\pi} \frac{ {(\bar\rho - \rv)}\,V} {\rl - \rv}}\;,
	  \label{eq:rem}
	\eea
	with the average density $\bar\rho=  (1/V) \int_{{\cal V}} \rho(\vect r)\,d\vect r$ in the box. For the lattice gas, one has to use
	the equilibrium profile $\req(\pmb{s})$ and replace the integral by a sum over all lattice points.
	The second definition is the so--called surface of tension which rests on a mechanical definition.
	First, one defines the surface tension $\gamma$ as an excess grand potential
	\bea
	  \gamma \ad = \Omega - ( - \pli \vd - \pv (V-\vd)) \;,  	
	  \label{eq:gamma}
	\eea
	and from the definition it is clear that the surface tension is not unique but depends on the choice of the
	droplet surface. Second, the radius $\Rs$ of the spherical drop surface of tension is fixed by the Laplace condition
	\bea
           \Delta p = \pli - \pv = \frac{2 \gamma}{\Rs}\;.
           \label{eq:rs}
	\eea
	It is useful to introduce the (radius-dependent) Tolman length by
	\bea
	   \delta(\Rs) = \Rem(\Rs) - \Rs \;.
	\eea
	
	From the Gibbs adsorption equation, one can derive a differential equation for $\gamma(\Rs)$, named the
	Gibbs--Tolman--Koenig--Buff (GTKB) equation \cite{Tolman1949}:
	\bea
          \frac{d\ln\gamma}{d\ln \Rs} &=& 
                \frac{\frac{2\delta}{\Rs}\left[ 1+\frac{\delta}{\Rs}+\frac{1}{3}\left( \frac{\delta}{\Rs}\right) ^2\right]}{1+\frac{2\delta}{\Rs}\left[ 1+\frac{\delta}{\Rs}+\frac{1}{3}\left( \frac{\delta}{\Rs}\right) ^2\right]} 
		\label{eq:GTKB}
	\eea
	A solution for $\gamma(\Rs)$ require knowledge of the function $\delta(\Rs)$, so the GTKB equation should be viewed
	as the consistency relation between these two functions in the first place. 

	We proceed by an {\em ansatz} for an expansion of $\gamma(\Rs)$, valid for large radii \cite{Troester2017}:
	\bea
		\frac{\gamma(\Rs)}{\gamma_\infty} = 1 - \frac{2\delta_\infty}{\Rs} + 
		\frac{A}{\beta \gamma_\infty}\, \frac{\ln (\Rs/a) }{\Rs^2} + \frac{B}{\beta \gamma_\infty}\,\frac{1}{\Rs^2} +\dots 
	\eea
	Here, $\gamma_\infty = \gamma(\Rs \to \infty)$ and likewise $\delta_\infty = \delta(\Rs \to \infty)$.
	Furthermore, $\beta=1/(\kt)$ is the inverse temperature and $a$ is a microscopic length, characteristic for the
	system (for the lattice gas, $a$ is simply the unit cell side length and is put to 1).
	This {\em ansatz} combines a simple $1/\Rs$--expansion with capillary--wave droplet shape fluctuations \cite{Wallace1980}, here
	the constant $A= -7/(12\pi) \approx -0.19$ is universal and can be obtained from integrating over droplet interface fluctuations
	using a standard Landau--Ginzburg Hamiltonian \cite{Prestipino2012,Prestipino2013,Prestipino2014}. Note that the exact value of $a$ is unimportant,
	using a different value $a'$ just results in a shift in the nonuniversal constant $B$ by $A \ln(a'/a)$.
	This expansion of $\gamma(\Rs)$ is consistent with the GTKB equation if the following expansion for the
	Tolman length holds:
	\bea
		\delta(\Rs) &=& \delta_\infty - \frac{A}{\beta \gamma_\infty}\, \frac{\ln (\Rs/a) }{\Rs}
		 		+ \left( \frac{\frac{1}{2}A -B}{\beta \gamma_\infty} +  3 \delta_\infty^2 \right)\,\frac{1}{\Rs} + \dots
	\eea
	In the lattice gas there is a symmetry between the liquid and the gas phase. This entails a vanishing Tolman length $\delta_\infty$.
	If furthermore there {were} no logarithmic term in the expansion, the  coefficient for the linear term $\propto -1/\Rs$ 
	in the Tolman length and the quadratic term $\propto 1/\Rs^2$ in the surface tension {would be} equal. This will be important below.

	\section{Results}
	\label{sec:results}

	\begin{figure}
	  \centerline{\includegraphics[width=8cm]{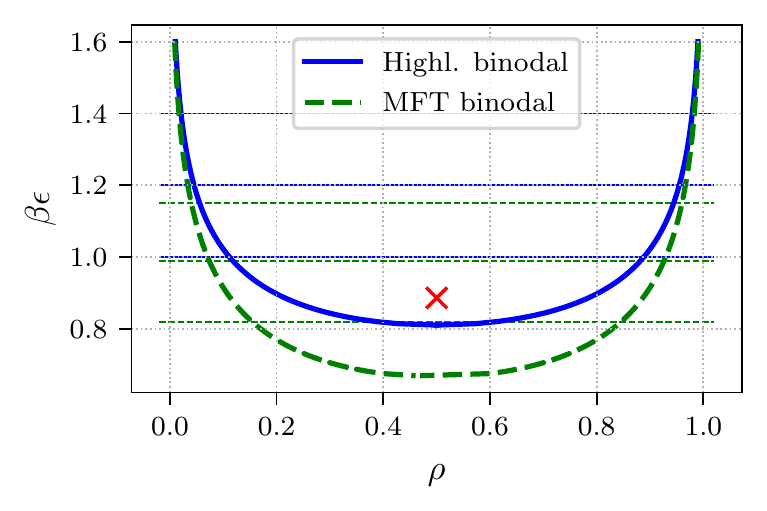}}
	  \caption{Binodal for the Highlander functional and the standard mean--field functional. For the Highlander functional,
	  we investigated droplets at the three temperatures $\beta\epsilon=1.0$, $1.2$ and $1.4$ (horizontal blue solid lines) and for the
	  mean--field functional we chose the three temperatures $\beta\epsilon=0.82$, $0.99$ and $1.15$ (horizontal green dashed lines). The red cross
	  marks the quasi--exact critical point from simulations \cite{Landau2018} $\beta_\text{c,sim} \epsilon\approx 0.887$.}
	  \label{fig:phasediag}
	\end{figure}

We investigated droplet formation for the Highlander functional at three different temperatures, characterized by
$\beta\epsilon=1.0$, $1.2$ and $1.4$. These are not close to the critical point (see Fig.~\ref{fig:phasediag} for
a phase diagram) but are also higher than the roughening temperature $\beta\epsilon_\text{rough} \approx 1.64$ \cite{Mon1990}. For comparison, we also calculated droplets using the standard mean--field functional at 
the three three temperatures $\beta\epsilon=0.82$, $0.99$ and $1.15$ which are roughly at the same relative distance to the
mean field critical point as the Highlander temperatures are from the Bethe--Peierls critical point. 

\subsection{$\mu$--equation of state and droplet solutions}

	\begin{figure}
	  \center
	  \includegraphics{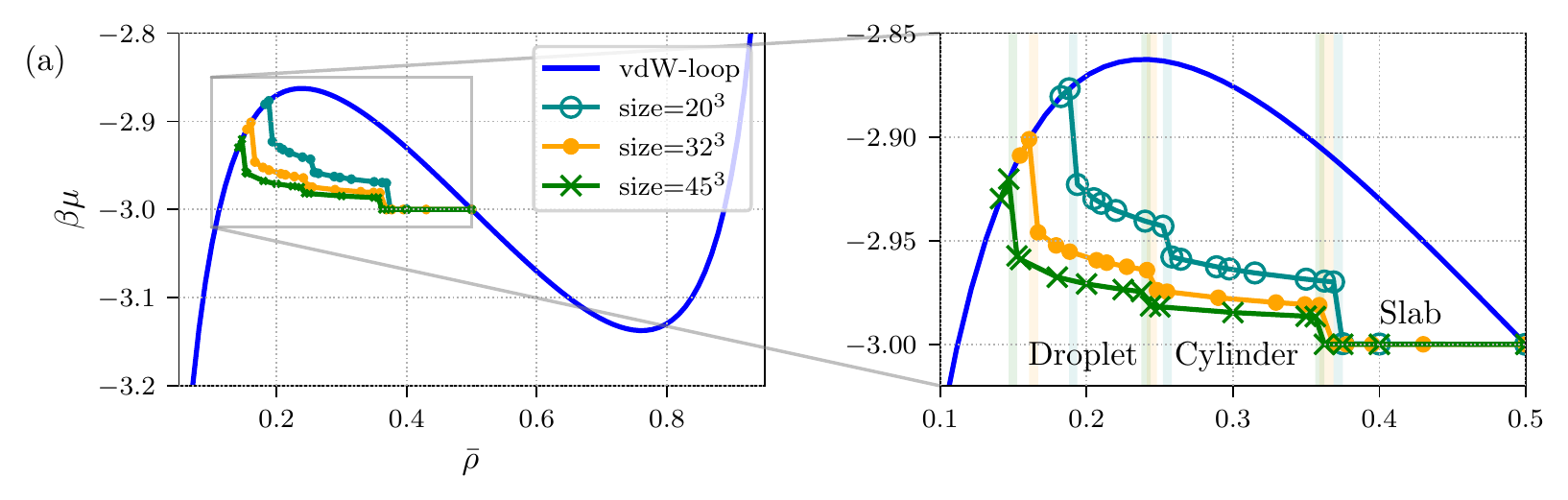}
	  \includegraphics{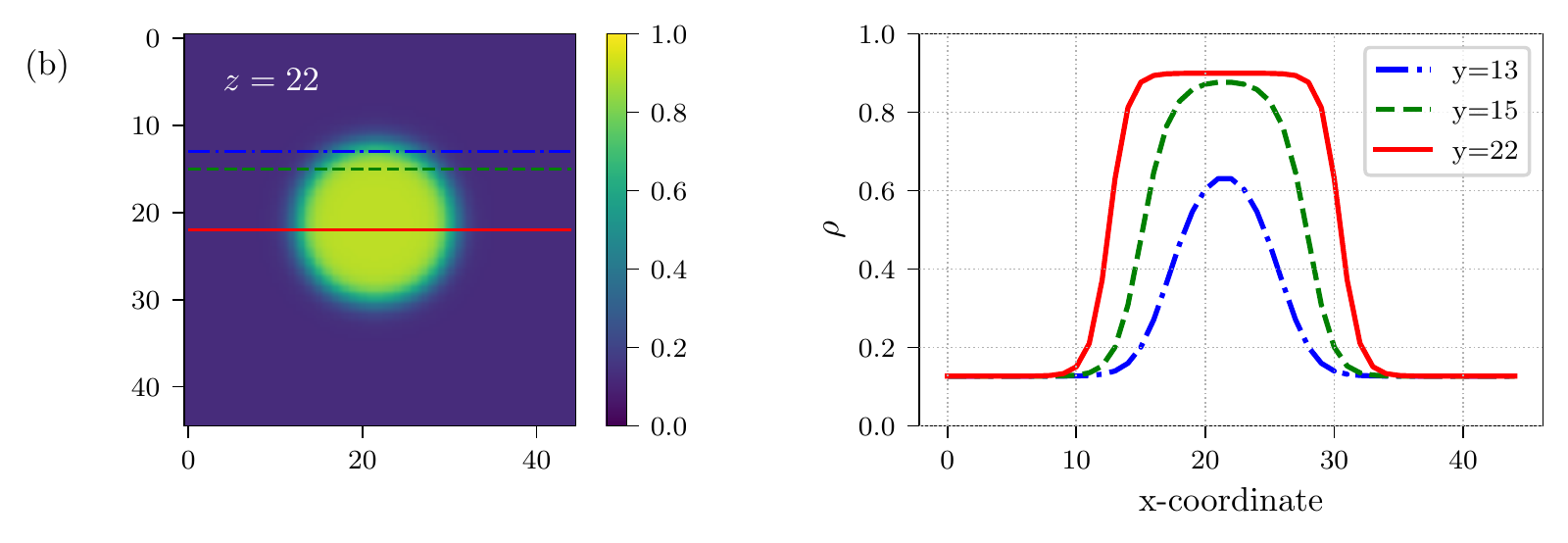}
	  \caption{(a) $\mu$--equation of state for the Highlander functional at $\beta\epsilon=1.0$ for three system sizes.  (b) Density profile for a stable droplet in the midplane of the box at $\bar\rho=0.155$ for system size $45^3$ with one-dimensional cuts.}
	  \label{fig:hi_mu_1.0}
	\end{figure}

For a finite system, one can obtain solutions for the full range of average densities $\bar\rho$ between the coexistence
densities $\rhovc$ and $\rholc$. These define a curve $\mu(\bar\rho)$ (different for each system size) which we call ``$\mu$--equation of state''. 

At the highest temperature $\beta\epsilon=1.0$ for the Highlander functional, $\mu(\bar\rho)$ shows the typical behavior  
known from previous work \cite{Schrader2009,Schmitz2013}, see Fig.~\ref{fig:hi_mu_1.0}(a). For the average density $\bar\rho$ slightly larger than $\rhovc$,
the system stays homogeneous and follows the bulk $\mu$--equation of state (shown as vdW--loop in  Fig.~\ref{fig:hi_mu_1.0}(a)).
Increasing $\bar\rho$, a transition to a droplet state occurs, characterized by a sharp downward jump in $\mu$. A second sharp jump
occurs which marks the transition to a cylinder state, and the third sharp jump marks the transition to a state with two planar interfaces (slab state). In Fig.~\ref{fig:hi_mu_1.0}(b) we show an example for a 2D cut through the midplane of the box of the 3D droplet density profile, where the system size is $45^3$ and the average density is $\bar\rho=0.155$. Evidently the droplet has a smooth interface with a width of a few lattice units.  

	\begin{figure}
	  \center
	  \includegraphics{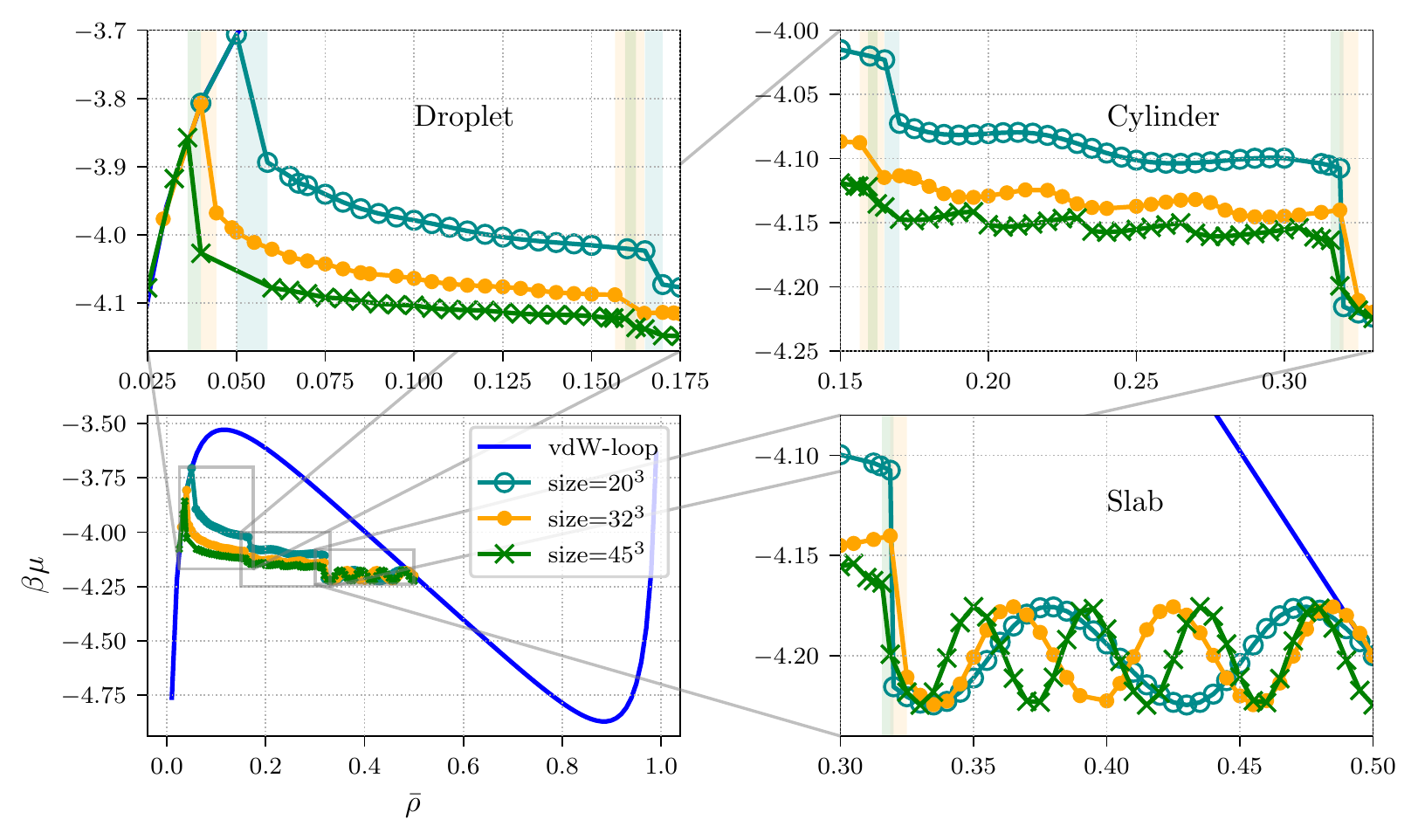}
	  \caption{$\mu$--equation of state for the Highlander functional at $\beta\epsilon=1.4$ for three system sizes. Enlargements show $\mu(\bar\rho)$ for the droplet, cylinder and slab portions of the $\mu$--equation of state.  }
	  \label{fig:hi_mu_1.4}
	\end{figure}

Lower temperatures introduce changes in $\mu(\bar\rho)$. First, oscillations appear in the slab portion of $\mu(\bar\rho)$ and by lowering the temperature further, kinky oscillations appear in the cylinder portion of $\mu(\bar\rho)$. This is shown in
Fig.~\ref{fig:hi_mu_1.4} where the $\mu$--equation of state is shown for $\beta\epsilon=1.4$ with suitable enlargements of the different
portions. The droplet portion
of $\mu(\bar\rho)$ appears to be smooth (non--oscillatory) but we will show below that some unusual behavior is seen in the
extracted droplet radii. 

We discuss first the oscillations in the slab portion. There we have two planar interfaces which form
under the constraint of a fixed average density. In Ref.~\cite{Maeritz2021} we had investigated one planar interface with such a constraint which
simply translates into a constraint for the equimolar position of the interface.  
In the lattice gas model, the properties of the interface are a priori only invariant upon discrete shifts
of the interface. Suppose that through the average density constraint one tries to put a few additional particles into
a system with a free, equilibrium interface. These additional particles can be accomodated by displacing the interface towards the vapor phase or they are added in the bulk, i.e. they change the bulk densities. In general, we find that both mechanisms 
occur,  and we had found solutions  at 
chemical potentials $\mu \neq \mucoex$, i.e., these are solutions for a planar interface off-coexistence. Necessarily these
solutions are found with a periodicity of 1 in the  equimolar position of the interface and since the interface position is proportional to the average density in the slab system, the oscillations
in $\mu(\bar\rho)$ result. 

These oscillations in the slab portion then also translate to the cylinder portion. 
It is reasonable to assume that there are certain relaxed cylindrical droplet states whose radii differ by discrete amounts. If one adds more particles to such a state, they may add to the bulk density away from the cylinder droplet (thus changing $\mu$) or increase the cylinder radius. Depending on which process dominates, changes in $\mu$ with $\bar\rho$ are steeper and lead to the spiky oscillations.

	\begin{figure}
	\centering
	  \includegraphics{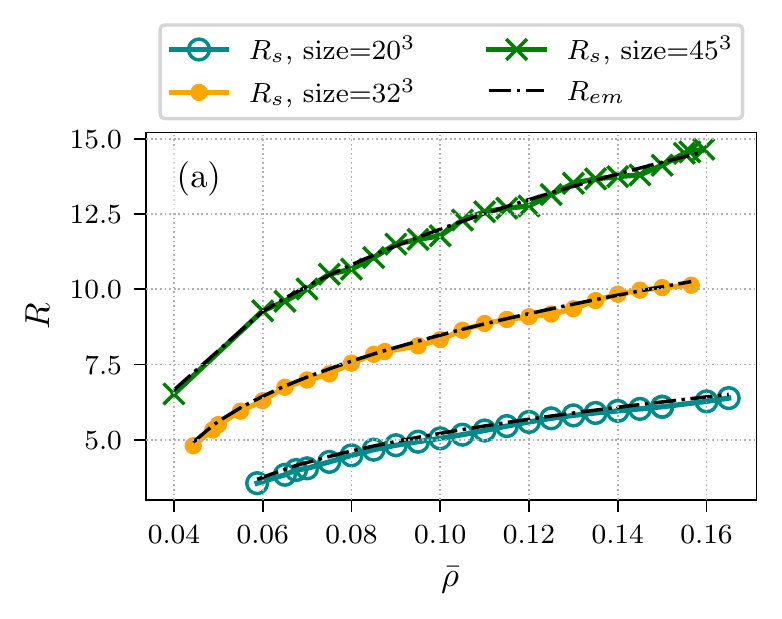}
	  \includegraphics{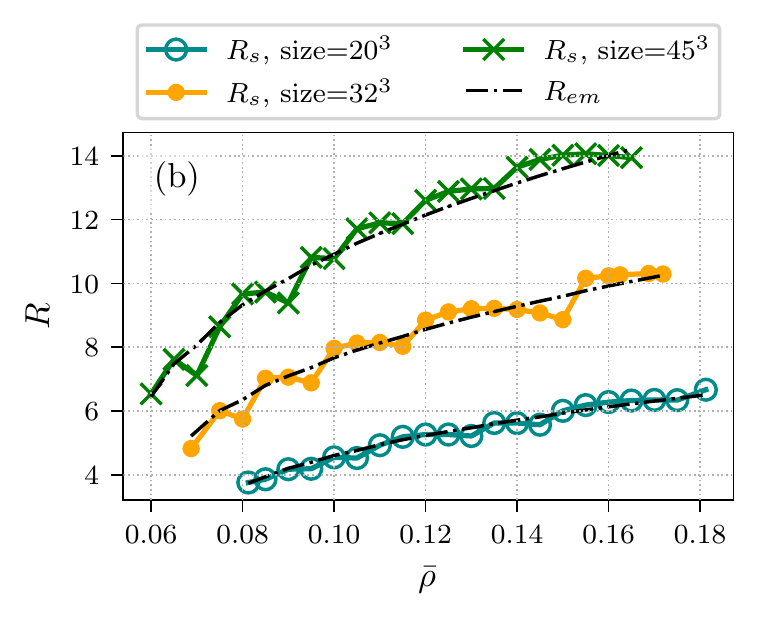}
	  \caption{ Equimolar droplet radius $\Rem$ and radius $\Rs$ of the surface of tension for (a) the Highlander functional and $\beta\epsilon=1.4$ and (b) the mean--field functional and $\beta\epsilon=1.15$.    }
	  \label{fig:rers}
	\end{figure}

In the droplet portion of the $\mu$--equation of state such oscillations are not obvious at first sight. From the density profiles we extract the equimolar radius through Eq.~\eqref{eq:rem} and the radius of the surface of tension through Eq.~\eqref{eq:rs}. In Fig.~\ref{fig:rers}(a), these
are shown for $\beta\epsilon=1.4$ (Highlander functional). The equimolar radius is a smooth function of the average density $\bar\rho$ but the radius $\Rs$ shows undulations, especially for the largest system which accomodates the largest droplets. The radius of the surface of tension is closest connected to the classical sharp surface and also closely related to the location where the density profile changes most strongly. Thus we see that the droplet radius increase upon increasing $\bar\rho$ proceeds unsteadily. 
For the simple mean--field functional, this effect is much stronger. In Fig.~\ref{fig:rers}(b), both radii are plotted for $\beta\epsilon=1.15$ and the mean--field functional. For the second--largest system, $\Rs$ changes with steps of roughly 1 upon increasing $\bar\rho$, and for the largest system the steps are occasionally larger. This can be viewed as a manifestation of a {\em droplet layering transition}. In App.~\ref{sec:appB}, we give strong evidence that in 2D and using the simple mean--field functional, the droplet layering transition is indeed sharp. The Highlander functional (which includes more `fluctuations'') shows this
transition only washed--out. This phenomenology is qualitatively similar to layering transitions at flat surfaces, e.g. in the continuum AO model where DFT predicts a whole sequence of discrete jumps \cite{Brader2003} in the adsorption but simulations resolve only the first three jumps \cite{Dijkstra2002}.  

\subsection{Asymptotic behavior of $\gamma(\Rs)$ and $\delta(\Rs)$ }

For the discussion of the radius--dependent behavior of the surface tension $\gamma(\Rs)$ (defined
through Eqs.~\eqref{eq:gamma} and \eqref{eq:rs}) and the radius--dependent Tolman length
$\delta(\Rs) = \Rem(\Rs) - \Rs$ we employ a fit of $\gamma(\Rs)$ to the following asymptotic form:
\begin{equation}
    \beta\gamma(\Rs) =
    \beta\tilde{\gamma}_\infty
    + A\, \frac{\ln \Rs }{\Rs^2}
    + B\,\frac{1}{\Rs^2} \;.
    \label{eq:fit gamma}
\end{equation}
In fit 1, the universal value of $A=-7/(12\pi) \approx-0.19$ is used, and only $B$ and $\tg$ are fitted. In fit 2, all three constants $A$, $B$ and $\tg$ are fitted and in fit 3 we put $A=0$ and again only $B$ and $\tg$ are fitted. The fit coefficients are used in the equation for the Tolman length 
\begin{equation}
    \delta(\Rs)= - \frac{A}{\beta\tilde{\gamma}_\infty}\frac{\ln(\Rs)}{\Rs}+\frac{\frac{A}{2}-B}{\beta\tilde{\gamma}_\infty}\frac{1}{\Rs}
    \label{eq:tolman}
\end{equation}
which follows through consistency with the GTKB equation \eqref{eq:GTKB} and checked against the numerical results for $\delta(\Rs)$.

    \begin{figure}
	  \centerline{\includegraphics[width=7.5cm]{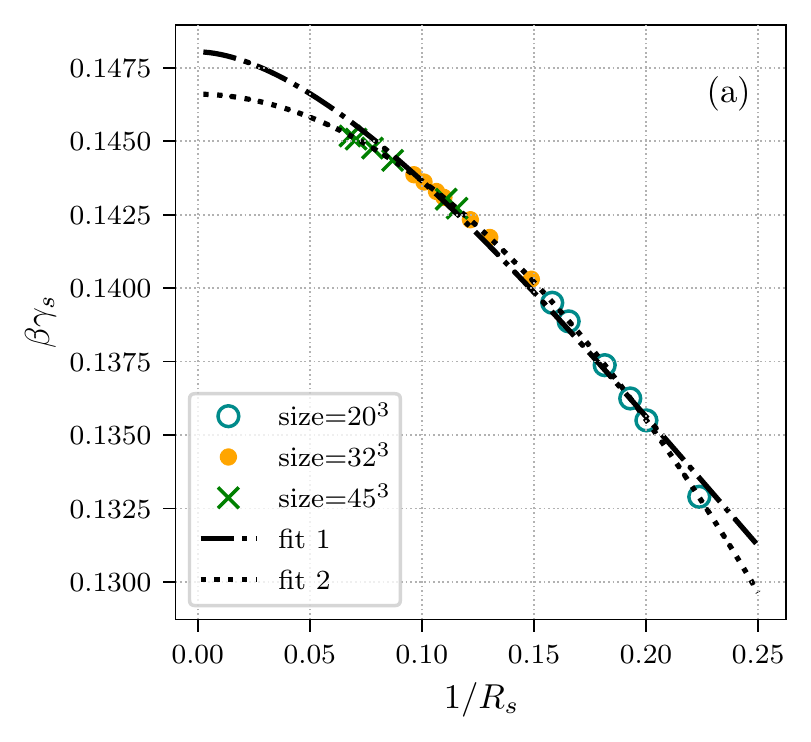}
	  \includegraphics[width=7.5cm]{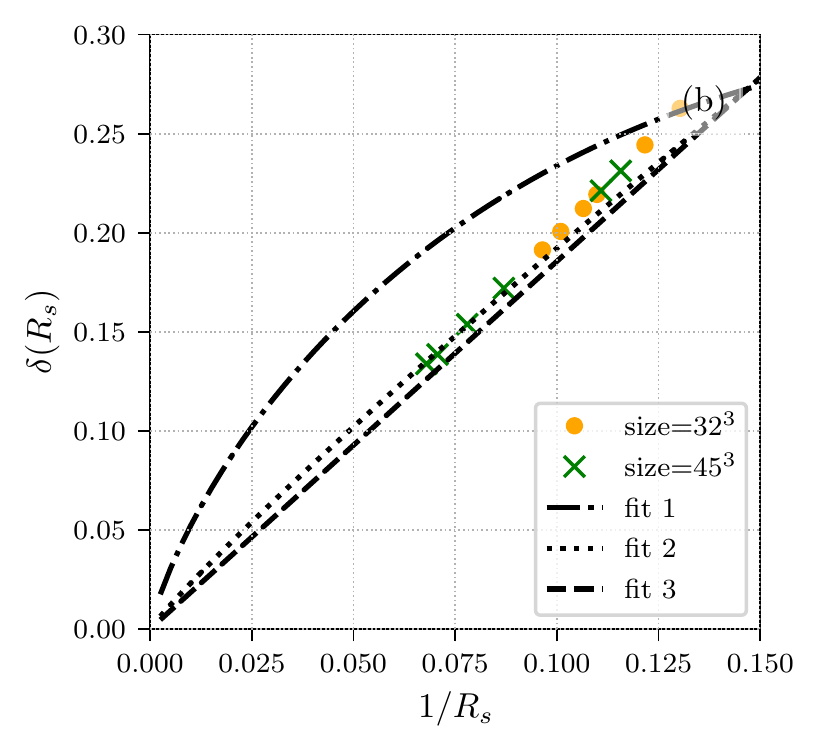}}
	  \caption{ Surface tension $\gamma(\Rs)$ (a) and Tolman-length (b) vs. $1/\Rs$ for the Highlander functional at $\beta\epsilon=1.0$. The fits in (a) are according to Eq.~(\ref{eq:fit gamma}) with $A$ considered universal (fit~1) or a fit parameter (fit~2) or $A=0$ (fit~3). The resulting fit parameters are shown in Tab.~\ref{tab:fit par}. The dotted, dash-dotted and dashed lines in figure (b) are according to the GTKB--consistent Eq.~(\ref{eq:tolman}) with the parameters taken from fit~1, 2 and 3 from (a). In contrast to (b), fit~2 and 3 are visually indistinguishable in (a), therefore only fit~2 is shown in (a). 
	  }
	  \label{fig:gamma/delta@epsi=1}
	\end{figure}

    \begin{figure}
	  \centerline{\includegraphics[width=7.5cm]{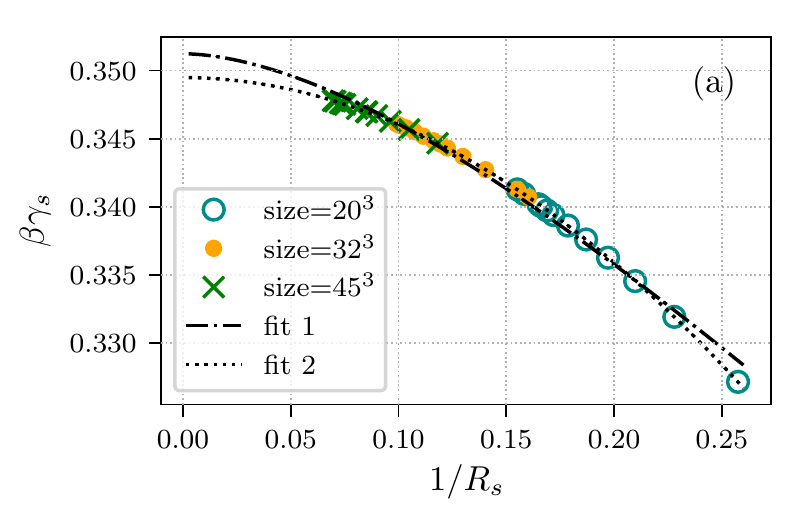}
	  \includegraphics[width=7.5cm]{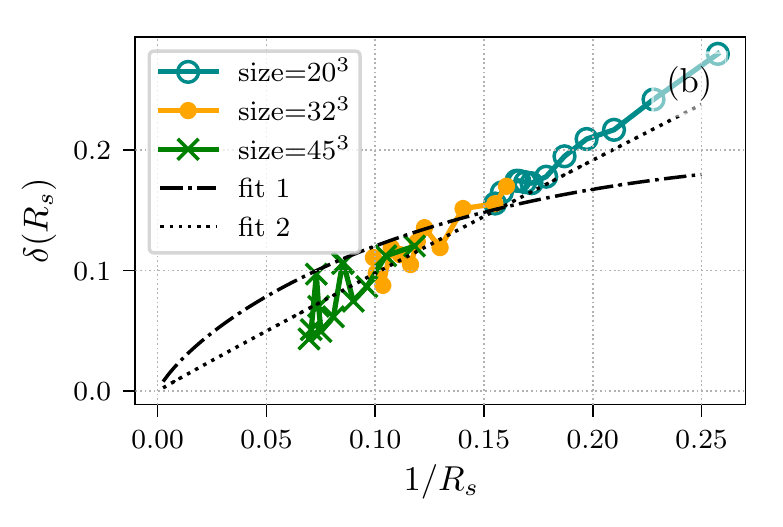}}
	  \centerline{\includegraphics[width=7.5cm]{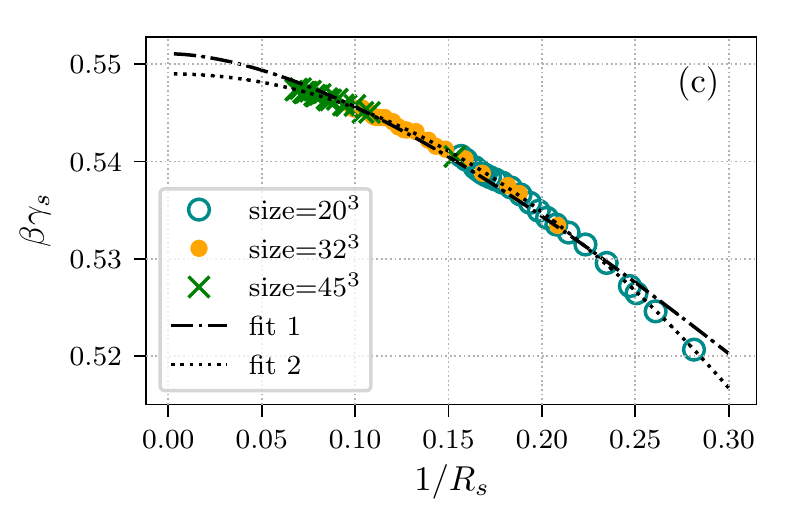}
	  \includegraphics[width=7.5cm]{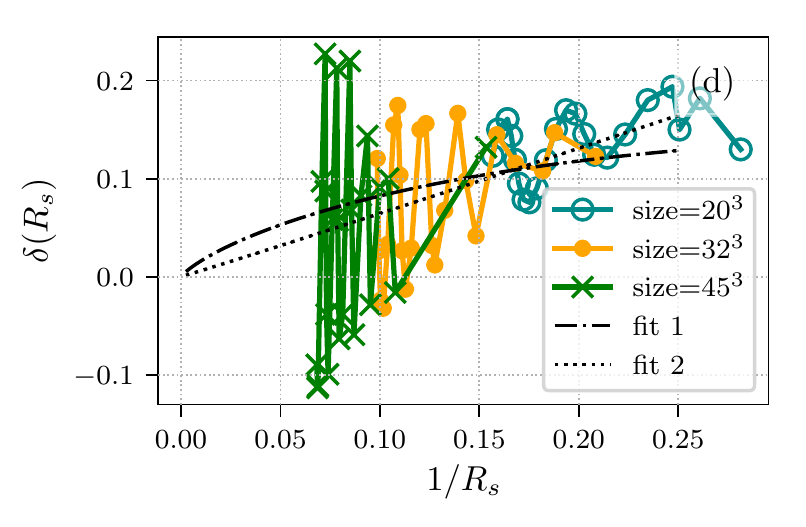}}
	  \caption{ Surface tension $\gamma(\Rs)$ (a,c) and Tolman-length (b,d) vs. $1/\Rs$ for the Highlander functional at $\beta\epsilon=1.2$ and $1.4$. For information on the fits see the caption of Fig.~\ref{fig:gamma/delta@epsi=1}. Fits~2 and 3 are indistinguishable in (a,c)  and their difference
	  is not very illuminating in (b,d) owing to the oscillatory numerical data, therefore only fits~1 and 2 are shown.}
	  \label{fig:gamma/delta;rest}
	\end{figure}
	
	Due to the discreteness of the lattice model, the planar interface tension $\gamma_\infty$ (weakly) depends on the orientation of the interface \cite{Bittner2009}. For a droplet, the surface tension $\tg$ in the limit of $\Rs\to\infty$ will be an average of the planar surface tension with different orientations. For simplicity, we treat $\tg$ as a fit parameter and discuss the fit result in the light of simulation and our own results for the anisotropy of $\gamma_\infty$.

    The results for the fit coefficients are found in Tab.~\ref{tab:fit par} and the comparison of the fits to the numerical data for the Highlander functional at  $\beta\epsilon=1.0$ is shown in Fig.~\ref{fig:gamma/delta@epsi=1} and at $\beta\epsilon=1.2$ and $1.4$ in Fig.~\ref{fig:gamma/delta;rest}. For all three temperatures, the data for $\gamma(\Rs)$ are smooth while the data for $\delta(\Rs)$ are only smooth for $\beta\epsilon=1.0$. For the two lower temperatures the wiggly behavior
    of $\Rs$ (see Fig.~\ref{fig:rers}(a)) leads to the spikes in $\delta(\Rs)$. For all temperatures, fit~1 with $A$ fixed to the universal constant only works moderately well for the surface tension. The ensuing asymptotic form for $\delta(\Rs)$ does not match the numerical data. Fit~2 (with $A$ free) and fit~3 (with $A=0$) work equally well for $\gamma(\Rs)$ and are not distinguishable on the plots. For $\beta\epsilon=1.0$, fit 2 matches the data points for
    the Tolman length for larger radii rather well, whereas there is a discrepancy for fit~3 which ignores the logarithmic term (see Fig.~\ref{fig:gamma/delta@epsi=1}(b)). For the two lower temperatures, the spiky behavior of $\delta(\Rs)$ does not allow for a discrimination in quality between fit~2 and fit~3. 
    
    It is perhaps not surprising that the asymptotic behavior of $\gamma(\Rs)$ and $\delta(\Rs)$ from DFT does not exhibit the universal logarithmic term. E.g., for the continuum Lennard--Jones fluid, expressions in an $1/R_s$--expansion had been derived for a very simple DFT model without the logarithmic term \cite{Giessen1998}. The analytic expansion for the full theory and approximate DFT's can be different but the overall description of $\gamma(\Rs)$ can nevertheless be good (if the underlying functional is). For the lattice gas droplets in Highlander DFT, the
    corrections to the surface tension are dominated by the $1/\Rs^2$--term but it is interesting to note that this term as the leading term is not fully consistent with the behavior of $\delta(\Rs)$ according to the GTKB equation, and a small logarithmic term is needed to restore the consistency. 

	\renewcommand{\arraystretch}{1.4}
	\begin{table}[]
	    \centering 
	    \begin{tabular}{C{1cm} c C{3cm} C{3cm} C{3.7cm}} \hline \hline 
	        & & $A$ & $B$ & $\beta\tilde{\gamma}_\infty$ \\ \hline
	        \multirow{3}{*}{\rotatebox{90}{$\beta\epsilon=1.0$}} & fit 1 & $-\frac{7}{12 \pi}$ & $-0.011\pm0.011$ & $0.14804\pm0.00026$\\
	        & fit 2 & $-0.025\pm0.004$ & $-0.236\pm0.005$ & $0.14660\pm0.00003$ \\
	        & fit 3 & $0$ & $-0.272\pm0.002$ & $0.14637\pm0.00004$ \\
	        \hline
	        \multirow{3}{*}{\rotatebox{90}{$\beta\epsilon=1.2$}} & fit 1 & $-\frac{7}{12 \pi}$ & $-0.088\pm0.011$ & $0.35123\pm0.00026$\\
	        & fit 2 & $-0.010\pm0.002$ & $-0.324\pm0.003$ & $0.34949\pm0.00002$ \\
	        & fit 3 & $0$ & $-0.337\pm0.001$ & $0.34939\pm0.00001$ \\
	        \hline
	        \multirow{3}{*}{\rotatebox{90}{$\beta\epsilon=1.4$}} & fit 1 & $-\frac{7}{12 \pi}$ & $-0.119\pm0.009$ & $0.55108\pm0.00024$\\
	        & fit 2 & $0.008\pm0.012$ & $-0.369\pm0.015$ & $0.54901\pm0.00014$ \\
	        & fit 3 & $0$ & $-0.358\pm0.002$ & $0.54910\pm0.00005$ \\
	    \hline \hline
	    \end{tabular}
	    \caption{Coefficients $A$, $B$ and $\beta \tg$ from the three fits of the Highlander numerical results for the surface tension to Eq.~(\ref{eq:fit gamma}).}
	    \label{tab:fit par}
	\end{table}
	
	\begin{table}[]
	    \centering 
	    \begin{tabular}{C{1cm} c C{3cm} C{3cm} C{3.7cm}} \hline \hline 
	        & & $A$ & $B$ & $\beta\tilde{\gamma}_\infty$ \\ \hline
	        \multirow{3}{*}{\rotatebox{90}{$\beta\epsilon=0.82$}} & fit 1 & $-\frac{7}{12 \pi}$ & $0.132\pm0.016$ & $0.10655\pm0.00031$ \\
	        & fit 2 & $-0.008\pm0.002$ & $-0.12\pm0.002$ & $0.10499\pm0.00001$ \\
	        & fit 3 & $0$ & $-0.131\pm0.001$ & $0.10493\pm0.00001$ \\
	        \hline
	        \multirow{3}{*}{\rotatebox{90}{$\beta\epsilon=0.99$}} & fit 1 & $-\frac{7}{12 \pi}$ & $0.087\pm0.008$ & $0.27152\pm0.00018$\\
	        & fit 2 & $-0.009\pm0.02$ & $-0.158\pm0.028$ & $0.26993\pm0.00019$ \\
	        & fit 3 & $0$ & $-0.171\pm0.003$ & $0.26985\pm0.00007$\\
	        \hline
	        \multirow{3}{*}{\rotatebox{90}{$\beta\epsilon=1.15$}} & fit 1 & $-\frac{7}{12 \pi}$ & $0.046\pm0.012$ & $0.43086\pm0.00032$\\
	        & fit 2 & $-0.055\pm0.063$ & $-0.124\pm0.083$ & $0.42949\pm0.00072$ \\
	        & fit 3 & $0$ & $-0.196\pm0.011$ & $0.42891\pm0.00029$ \\
	    \hline \hline
	    \end{tabular}
	    \caption{Coefficients $A$, $B$ and $\beta \tg$ from the three fits of the mean field numerical results for the surface tension to Eq.~(\ref{eq:fit gamma}).}
	    \label{tab:fit par mf}
	\end{table}

For fits~2 and 3, the asymptotic surface tension $\beta\tg$ lies between the surface tensions
$\beta\gamma_\infty$ for the interface in [100] orientation and in [111] orientation, see
Tab.~\ref{tab:g_aniso}. For fit~1, the value is outside or very close to $\beta\gamma_\infty [111]$. The Highlander values for the anisotropy are reasonably close to values from simulations \cite{Bittner2009} if one compares for temperatures with the same $\beta_\text{c}/\beta$. 

	\begin{table}[]
	    \centering 
	    \begin{tabular}{C{2cm} c C{3cm} C{3cm} C{3cm}} \hline \hline 
	    $\beta\epsilon$ & $\beta\gamma_\infty[100]$ & $\beta\gamma_\infty[110]$ & $\beta\gamma_\infty[111]$ \\ \hline
	    1.0 & 0.145 & 	0.147	 & 0.147 \\
        1.2 & 0.342 & 0.351 &	0.354 \\
        1.4 & 0.523 & 	0.554 & 0.561 \\	
        \hline \hline
	    \end{tabular}
	    \caption{Planar surface tension from slab configurations for different orientations of the interface for the Highlander functional, using
	    a constraint $\bar{\rho}=0.5$ for the average density in the system. 
	    }
	    \label{tab:g_aniso}
	\end{table}

For the droplet surface tension from the simple mean-field functional, the fit results are reported
in Tab.~\ref{tab:fit par mf}. For the two higher temperatures ($\beta\epsilon=0.82$ and 0.99), the main radius dependence
in $\gamma(\Rs)$ is in the $1/\Rs^2$ term, note that the associated coefficient $B$ is only about half of the $B$ value from the
Highlander functional. Thus, the improvement in the functional leads to a significantly larger radius dependence of the surface tension. For the lowest temperature ($\beta\epsilon=1.15$) the surface tension shows spikes due to the layering transition
(see Fig.~\ref{fig:rers}(b) for the manifestation in the radii) and the fits are not particularly meaningful.

	

\section{Summary and conclusions}
\label{sec:summary}

In this paper, we have applied a recently developed density functional for the lattice gas (Ising model) to the problem of droplet states in finite systems. We have found the sequence of droplets to cylinders to planar slabs upon increasing the average density $\bar\rho$ in the system similar to previous simulation studies. Owing to the discreteness of the lattice, we have seen additional effects in the state curve $\mu(\bar\rho)$ (the $\mu$--equation of state). Upon lowering the temperature away from the critical temperature, we first find oscillations in $\mu(\bar\rho)$ in the slab portion. When decreasing the temperature further, spiky undulations in $\mu(\bar\rho)$ in the cylinder portion are seen and in the droplet region an undulatory behavior of $\Rs(\bar\rho)$ (radius of the surface of tension) is found. We could relate this behavior in the cylinder and droplet region to washed--out layering transitions at the surface of liquid cylinders and droplets. The analysis of the large--radius behaviour of the surface tension $\gamma(\Rs)$ gave a dominant contribution $\propto 1/\Rs^2$, although the consistency of $\gamma(\Rs)$ with the asymptotic behavior of $\delta(\Rs)$ (the radius--dependent Tolman length) through the GTKB equation seems to suggest a weak logarithmic contribution $\propto \ln\Rs/\Rs^2$ in $\gamma(\Rs)$. The coefficient of this logarithmic term is smaller than a universal value derived with field--theoretic methods.

We remark that Ref.~\cite{Tovbin2010} utilizes a Bethe--Peierls approach for the lattice gas to calculate numerical solutions for droplets. However, further intermediate approximations are made in Ref.~\cite{Tovbin2010} (assumption of radially symmetric density profiles, assumption of
an effective distorted lattice at the droplet interface) such that a comparison is difficult.
While we find a reduction of the surface tension with decreasing radius for all investigated 
temperatures, an increase is found in Ref.~\cite{Tovbin2010} for low temperatures. 

Although the lattice gas (Ising model) has been studied thoroughly with simulations, no particular attention has been paid to the lattice discreteness effects observed here. It would be interesting to check for those in future simulations. In a wider perspective, it would also be desirable to develop DFT and simulation methods further for lattice or continuous systems to achieve agreement for the behavior of $\gamma(\Rs)$ over a wider range of droplet sizes. 

\begin{appendix}

\section{The excess free energy functional for the system of lattice gas particles and polymer clusters}
\label{app:fex}

The interactions between lattice gas particles and polymer clusters are characterized as follows:
Both the particles (with density $\rhoc(\pmb{s})$) and polymer clusters of a certain orientation $\alpha \in \{x,y,z\}$ 
(with density $\rhopca(\pmb{s})$) are separately hard lattice gases
(mutual exclusion of particles on one lattice site). Polymer clusters of different orientation do not interact
with each other. The lattice gas particles are excluded from the site of the polymer clusters and one additional site in $\alpha$--direction. Intrinsically, this is a nonadditive model, and the construction of the excess free energy
functional proceeds via lattice FMT \cite{Lafuente2002,Lafuente2004}. In 1D, this gives the exact functional \cite{Cuesta2005}.

The construction of lattice FMT functionals proceeds via the following iterative procedure \cite{Lafuente2004}. 
First, one finds a maximal set of 0D cavities.  A 0D cavity consists of a set of lattice points
for each species with the following property: If one particle of a certain species occupies one
of the points in the set, no other particle will fit in the cavity. The 0D cavity is maximal if
no further points can be added to the set.
The requirement on the excess functional is that it gives 
the exact 0D excess free energy for a density distribution, compatible with any such maximal cavity at an arbitrary location.
The exact 0D excess free energy of a cavity with occupation $\eta$ is given by $\beta^{-1} \pzd(\eta) $ with 
\bea
 \pzd(\eta) = \eta + (1-\eta) \ln (1-\eta) \;.
 \label{eq:Phi0D}
\eea
Second, the iterative procedure is started as follows: The excess free energy is a sum over
the 0D free energies of all such density distributions. However, when a specific cavity of the maximal set is evaluated with this trial 
functional, it will generate the correct 0D free energy plus some residual terms. All these residual terms are explicitly subtracted
in an updated excess free energy. Re--evaluation with a specific cavity may result in further residual terms which need to be subtracted again.
As shown in Ref.~\cite{Lafuente2005}, this procedure is guaranteed to terminate with no residual terms and thus the excess functional
has the desired property of giving the exact 0D free energy for any maximal cavity.
In our previous work, Ref.~\cite{Maeritz2021}, the construction is performed in detail for the lattice gas in 1D, 2D and 3D. 

In 3D, the iteration procedure leads to a set of 10 weighted densities:
\begin{align}
	m_1(\pmb{s})&:=\rho_{\text{pc},x}(\pmb{s})+\rho_\text{c}(\pmb{s}) \notag\\
	m_2(\pmb{s})&:=\rho_{\text{pc},x}(\pmb{s})+\rho_\text{c}(\pmb{s}+\hat{\pmb{e}}_x) \notag\\
	m_3(\pmb{s})&:=\rho_{\text{pc},y}(\pmb{s})+\rho_\text{c}(\pmb{s}) \notag\\
	m_4(\pmb{s})&:=\rho_{\text{pc},y}(\pmb{s})+\rho_\text{c}(\pmb{s}+\hat{\pmb{e}}_y) \notag\\
	m_5(\pmb{s})&:=\rho_{\text{pc},z}(\pmb{s})+\rho_\text{c}(\pmb{s}) \notag\\
	m_6(\pmb{s})&:=\rho_{\text{pc},z}(\pmb{s})+\rho_\text{c}(\pmb{s}+\hat{\pmb{e}}_z) \notag \\
    m_7(\pmb{s})&:=\rho_{\text{pc},x}(\pmb{s}) \notag \\ 
    m_8(\pmb{s})&:=\rho_{\text{pc},y}(\pmb{s}) \notag \\
	m_9(\pmb{s})&:=\rho_{\text{pc},z}(\pmb{s}) \notag \\
	m_{10}(\pmb{s})&:=\rho_{\text{c}}(\pmb{s})\qquad.
    \label{eq:n3D}
\end{align}
The excess free energy functional can be written with a local free energy density
\bea
  \beta \tFAO^\text{ex}=\sum_{\pmb{s}} \Phi_\text{AO,3D}(\pmb{s})
\eea
and this density $\Phi_\text{AO,3D}(\pmb{s})$ is given by
\begin{multline}
 \Phi_\text{AO,3D}(\pmb{s})=\pzd\left( m_1\right) +\pzd\left( m_2\right) +\pzd\left( m_3\right) + \pzd\left( m_{4}\right) + \pzd\left( m_{5}\right)+ \pzd\left( m_{6}\right)\\
- \pzd\left( m_{7}\right) - \pzd\left( m_{8}\right) - \pzd\left( m_{9}\right)- 5\,\pzd\left( m_{10}\right)
	\label{eq:Phi3Da} \;.
\end{multline}

\section{Droplet layering transition for the simple mean--field functional in 2D}
\label{sec:appB}

\begin{figure}
	  \center
	  \includegraphics[width=8cm]{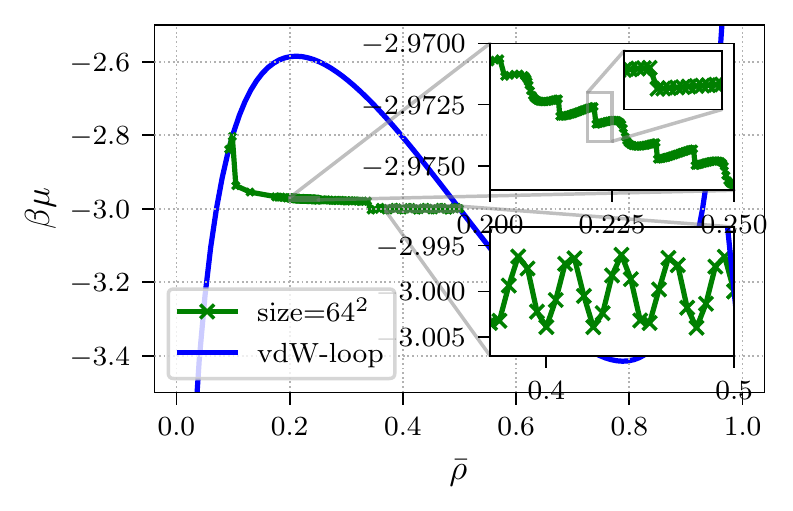}
	  \caption{$\mu$--equation of state for the mean--field functional in 2D at $\beta\epsilon=1.5$ for one system size $64^2$. Enlargements show $\mu(\bar\rho)$ for the circular droplet and slab portions of the $\mu$--equation of state. }
	  \label{fig:mf_mu2D_1.5}
	\end{figure}

\begin{figure}
	  \center
	  \includegraphics{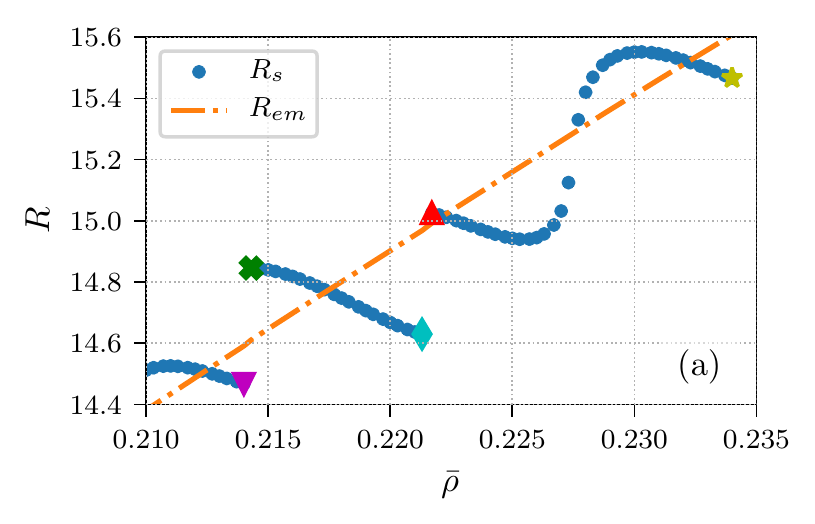}
	  \includegraphics{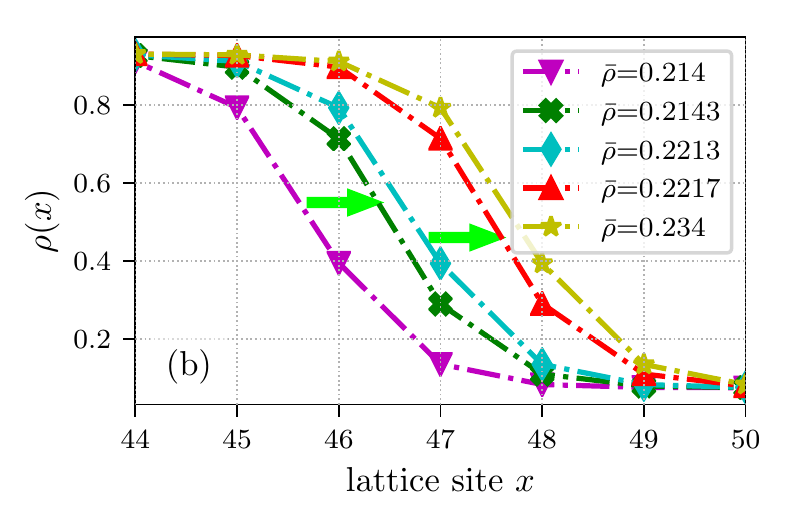}
	  \caption{Droplet solutions for the mean--field functional in 2D at $\beta\epsilon=1.5$. (a) $\Rem$ and $\Rs$ vs. $\bar\rho$, symbols mark states whose density profiles along the $x$--axis are shown in (b) using the same symbols. Arrows mark the two
	  discrete jumps seen in (a).}
	  \label{fig:mf_2D_1.5}
	\end{figure}

Stable droplets can also be generated in finite 2D boxes. Here we use the simple mean--field functional to investigate these droplets for a temperature of $\beta\epsilon=1.5$ (note that the mean--field critical point is at a value of 1.0). Fig.~\ref{fig:mf_mu2D_1.5} shows the $\mu$--equation of state which contains a circular droplet portion and a slab portion. In the droplet portion, two discrete jumps in $\mu$ are seen at $\bar\rho \approx 0.214$ and $\bar\rho \approx 0.221$. In Fig.~\ref{fig:mf_2D_1.5} density profiles of droplets are shown which correspond to states right to the left and to the right of each jump. E.g., the 
purple downward triangles and green crosses show the change in density profile at the first jump; the density profile changes by adding approximately one layer to the circular droplet. The same holds for the turquoise diamonds and red upward triangle which correspond to the second jump. So the two jumps in the chemical potential correspond to two layering transitions in the droplet profiles. For comparison, another density profile at a higher $\bar\rho=0.234$ (beyond the second jump at $\bar\rho=0.221$) is shown which shows only a small change compared to the profile at the second jump. 

In 2D, these droplet layering transitions are completely gone when using the Highlander functional, we checked this for temperatures down to $\beta\epsilon=3.0$. Accordingly we assume that they do not exist in simulations in 2D as well.

\end{appendix}

\FloatBarrier

\bibliographystyle{PRE}

\end{document}